\documentclass[12pt,english]{article}
\usepackage[latin1]{inputenc}
\usepackage{epsf,amssymb,amsmath,latexsym}
\usepackage{times}
\usepackage{tabularx}
\usepackage[dvips]{color,graphicx}
\makeatletter

\usepackage{babel}
\makeatother

\newcommand{\be}{\begin{equation}}
\newcommand{\ee}{\end{equation}}
\newcommand{\bea}{\begin{eqnarray}}
\newcommand{\eea}{\end{eqnarray}}
\newcommand{\ba}{\begin{array}}
\newcommand{\ea}{\end{array}}

\begin{document}


\begin{titlepage}
\rightline{DAMTP-2007-27} \rightline{\tt{arXiv:0704.3728 [hep-th]}}

\vfill
\vfill
\vfill
\vfill

\begin{center}
\baselineskip=16pt {\Large\bf  Quantum Effects in Black Holes }
\vskip 0.2cm {\Large\bf from the Schwarzschild Black String?}
\vskip 0.3cm {\large {\sl }} \vskip 10.mm {\bf Alessandro Fabbri$^1$} \vskip 1cm
{\small
Departamento de F\'{\i}sica
Te\'orica and IFIC, Universidad de Valencia-CSIC, \\
C. Dr. Moliner 50, Burjassot-46100, Valencia, Spain.}

\vskip 0.2cm {\large {\sl }} \vskip 10.mm {\bf Giovanni Paolo Procopio$^2$} \vskip 1cm
{\small
D.A.M.T.P., Centre for Mathematical Sciences, University of Cambridge,  \\
Wilberforce Road, Cambridge CB3 0WA, U.K.}
\end{center}
\vfill

\par
\begin{center}
{\bf ABSTRACT}
\end{center}
\begin{quote}

The holographic conjecture for black holes localized on a 3-brane in
Randall-Sundrum braneworld models RS2 predicts the existence of a
classical 5D time dependent solution dual to a 4D evaporating black
hole. After briefly reviewing recent criticism and presenting some
difficulties in the holographic interpretation of the
Gregory-Laflamme instability, we simulate some basic features of
such a solution by studying null geodesics of the Schwarzschild
black string, in particular those propagating nontrivially in the
bulk, and using holographic arguments.

\vfill
\vfill
\vfill
\vfill
\vfill

\vfill
\hrule width 5.cm
\vskip 2.mm
{\small
\noindent $^1$  afabbri@ific.uv.es\\
\noindent $^2$ g.p.procopio@damtp.cam.ac.uk
\\ }
\end{quote}
\end{titlepage}

\setcounter{equation}{0}

\section{Introduction}

Holography is an important tool to relate seemingly different
theories living in different spacetime dimensions. The $AdS$/CFT
correspondence \cite{adscft} adapted to the Randall-Sundrum
braneworld model RS2 \cite{Randall:1999vf} predicts that classical
bulk physics in $AdS_{5}$ is dual to a particular cut-off CFT
(namely ${\cal{N}}=4$ SU($N$) SYM in the planar limit) coupled to
gravity living on the brane. A concrete realization of this is
provided by the calculation of the correction to the Newtonian
potential on the brane, which can be performed in two very
different ways giving the same result. Studying linear
gravitational perturbations (5D gravitons) of the form $h_{\mu
\nu}=e^{ipx}H_{\mu \nu}(z)$, with $p^2=m^2$, around the
Randall-Sundrum vacuum
\begin{equation}\label{rsvac}
    ds^2=e^{-2k|z|}\left[\eta_{\mu \nu}dx^{\mu}dx^{\nu}\right]+dz^2
\end{equation}
where $k$ is related to the five-dimensional cosmological constant
by $k=\sqrt{-\Lambda_5/6}$, Garriga and Tanaka \cite{Garriga:1999yh}
found that the gravitational potential on the brane ($z=0$)
generated by a mass $M$ is
\begin{equation}\label{grap1}
    \phi (r) =\frac{M}{r}\left(1+\frac{2}{3}\frac{1}{k^2r^2}\right)
    \ ,
\end{equation}
for scales $r\gg 1/k$. The first, Newtonian, term is given by the
zero-mode ($m^2=0$) bound to the brane, while the second,
corrective, term is the contribution induced by the massive modes
($m^2\ne 0$) living in the bulk. On the other hand, from a pure 4D
perspective, Duff \cite{Duff:1974ud} found that the quantum
corrected Newtonian potential is given by
\begin{equation}\label{grap12}
    \phi (r) =\frac{M}{r}\left(1+\frac{\alpha \hbar }{r^2}\right) \
    ,
\end{equation}
where the coefficient  $\alpha=\frac{2N^2}{3\pi}$ depends on the
relevant CFT, $N^2$ counting the number of degrees of freedom. If
one combines this result with the holographic relation
\begin{equation}\label{holrel}
    \frac{1}{k^2}=\frac{\hbar N^2}{\pi},
\end{equation}
 one recovers
exactly \eqref{grap1} \cite{Duff:2000mt} (see also
\cite{Anderson:2004md}). The equivalence between the expressions
\eqref{grap1} and \eqref{grap12} shows that the quantum
corrections due to the CFT are classically given by the bulk
massive modes $m^2 \neq 0 $.

Application of holographic ideas beyond the linearized level, in
particular to the extreme case of black holes, has led to the
conjecture that for large masses ``black hole solutions localized on
the brane in the $AdS_{D+1}$ braneworld which are found by solving
the classical bulk equations in $AdS_{D+1}$ correspond to
quantum-corrected black holes in $D$ dimensions, rather than
classical ones'' \cite{conj}.

Evidence for this conjecture comes from the explicit solutions of
black holes localized on a 2-brane in $AdS_{4}$ \cite{Empa} and on a
1-brane in $BTZ$ \cite{Germani:2006ea}, but in the physically
relevant case of a black hole on a 3-brane in $AdS_5$ things are
much more complicated and no static solutions of this type have been
found yet \cite{bhbra}. The lack of a full 5D braneworld solution
giving a 4D static asymptotically flat black hole localized on a
3-brane is naturally explained by the holographic conjecture: a
quantum corrected 4D black hole cannot be static as it would
evaporate via the Hawking effect.


Recently, as a counterexample to this argument the RS2
Schwarzschild black string solution \cite{Chamblin:1999by}
 \begin{equation}\label{efv}
ds^2=e^{-2k|z|}\left[-(1-\frac{2M}{r})dt^2 +
\frac{dr^2}{(1-\frac{2M}{r})} + r^2d\Omega^2\right]+dz^2
\end{equation}
was considered in \cite{Fitzpatrick:2006cd}. Indeed, this solution
gives the classical Schwarzschild solution on the brane. To
justify the absence of quantum corrections in the dual theory, it
was speculated in \cite{Fitzpatrick:2006cd} that due to strong
coupling effects the number of asymptotic degrees of freedom is
drastically reduced and this would imply that the radiation
vanishes at leading order (where only terms which are $O(\hbar
N^2)$ survive).

The arguments used in \cite{Fitzpatrick:2006cd} are based on the
fact that on a sphere of radius $R$ and at large 't Hooft coupling
$\lambda$ the energy separation for weakly interacting states is
$\sim \frac{\lambda^{1/4}}{R}$
and thus the spectrum is lifted to infinite energy apart from the
$O(1)$ massless states dual to the supergravity modes of the
string. However, note that in the flat limit $R\to\infty$ the
above mass gap disappears and this is consistent with the
perturbative calculation (\ref{grap1}), which in the dual CFT
gives $O(N^2)$ results. The same happens in cosmology
\cite{Tanaka:2004ig}. It is difficult to believe that the number
of massless degrees of freedom is in general $O(1)$ except for the
cases where explicit verification is possible, i.e. flat and
cosmological branes. Another problem with the proposed holographic
interpretation of \eqref{efv} is to explain why, besides the
absence of a radiative term, all the quantum corrections (say, the
vacuum polarization terms) are actually suppressed.
Moreover, as already pointed out in \cite{Chamblin:1999by}, the
singularity at $r=0$ extends all the way from $z=0$ to $z=\infty$,
making the AdS horizon singular as well.\footnote{The authors of
\cite{Fitzpatrick:2006cd} have tried to remove this problem by
considering an additional brane. This in the dual theory on the
brane implies that the CFT is cut-off also in the IR.} Therefore,
such solution is likely not to have a counterpart in the dual CFT.

It has been suggested that the Gregory-Laflamme (GL) instability
\cite{GL}, \cite{GL2} of \eqref{efv} is dual to Hawking radiation in
the boundary theory \cite{Chamblin:2004vr}. We will show in section
2 that a quantitative comparison of the GL  brane perturbed metric
with that of an evaporating 4D black hole in the near-horizon region
shows some problematic points. We will then turn, in section 3, to
an interesting feature of the RS2 Schwarzschild black string which
we discovered by studying null geodesics and in particular those
propagating nontrivially in the bulk. Indeed, in the `geometrical
optics' approximation, 5D gravitational waves (in particular, bulk
massive modes) travel along such geodesics and these, in turn, might
contain ``seeds'' of quantum effects in the dual theory. The
holographic interpretation of our results, in section 4, leads us to
conjecture some crucial features of the actual 5D configuration dual
to an evaporating 4D black hole. This is done in section 5. Finally,
in section 6 we briefly state our conclusions.

\section{GL instability and Hawking radiation}

An important property of the RS2 black string 
comes from the analysis of its linear gravitational perturbations
which leads to the well-known Gregory-Laflamme instability.
In view
of the holographic black holes conjecture \cite{conj} it is
natural to wonder whether such classical instability corresponds,
in the boundary theory, to the quantum instability of the 4D
Schwarzschild spacetime via Hawking radiation.

Considering the metric \eqref{efv}, it was shown by Gregory
\cite{GL2} that the instability for the flat black string initially
discovered in \cite{GL} simply generalizes to this warped case.
Perturbations of the metric
\be
    g_{\mu \nu}\rightarrow g_{\mu \nu}+ \delta g_{\mu \nu}\ ,
\ee where
\be \delta g_{\mu \nu}= u_{m}(z) \hat g_{\mu \nu}\ ,\ee take
the following form in the $r\rightarrow 2M$ limit
\begin{align}
 \hat g^{tt} & \approx   (-\frac{1}{2} + 2M\Omega )
(1-\frac{2M}{r})^{-2+2M\Omega}
 e^{\Omega t} \\
\hat g^{rr} & \approx (-\frac{1}{2} +
2M\Omega)(1-\frac{2M}{r})^{2M\Omega}
 e^{\Omega t}  \\
\hat g^{tr} & \approx  - (-\frac{1}{2} + 2M\Omega )
 (1-\frac{2M}{r})^{-1+2M\Omega} e^{\Omega t} \ ,
\end{align}
while
\begin{equation}
u_{m}(z) = {\cal A}\, {\rm J}_2 \left ( {m\over k} e^{k|z|} \right)
- {\cal B}\, {\rm N}_2 \left ( {m\over k} e^{k|z|} \right )
\end{equation}
with the coefficients ${\cal A}$ and ${\cal B}$
satisfying  \be {\cal A}{\rm J}_1\left ( {m\over k} \right ) =
{\cal B} {\rm N}_1 \left ( {m\over k} \right ).\ee

The suitable time coordinate to parameterize the (future) horizon
is not $t$ but $v$, the ingoing Eddington-Finkelstein null
coordinate ($v=t+r+2M\ln\frac{r-2M}{2M}$). In terms of $v$ the
above perturbations take the form
\begin{align}
 \hat g_{tt} & \approx   (-\frac{1}{2} + 2M\Omega )
 e^{\Omega v} \\
\hat g_{rr} & \approx (-\frac{1}{2} +
2M\Omega)(1-\frac{2M}{r})^{-2}
 e^{\Omega v}  \\
\hat g_{tr} & \approx   (-\frac{1}{2} + 2M\Omega )
 (1-\frac{2M}{r})^{-1} e^{\Omega v}\ .
\end{align}
These vanish at the past horizon ($v=-\infty$).

With the change of coordinates $(t,r)\to\ (v,r)$ the perturbed
metric on the brane ($z=0$) along the future horizon takes the
ingoing Vaidya form
\be\label{vaidya}
ds^2=-(1-\frac{2m(v)}{r})dv^2+2dvdr +r^2d\Omega^2\ ,\ee with mass
function given by \be \label{massa} m(v)=M+M(-\frac{1}{2} +
2M\Omega )e^{\Omega v}u_m(0)\ .\ee

The numerical results show that for $M=1$
the most favored instability has $m_0 \simeq 0.2$ and $\Omega_0
\simeq
 0.05$, while for general $M$ we have $m_0\to \frac{m_0}{M}$ and
 $\Omega_0\to
 \frac{\Omega_0}{M}$.
 In the large $M$ (i.e. $m\ll k$) limit the leading order term in
 $u_m(z)$ is
 \be
 u_m(z)= {\cal A}\left[ {\rm J}_2 \left( {m\over k} e^{k|z|} \right) -
 \frac{{\rm J}_1\left( {m\over k} \right)}{ {\rm N}_1 \left( {m\over k} \right)}
 {\rm N}_2 \left( {m\over k} e^{k|z|} \right)\right]\sim {\cal A} e^{-2k|z|}\ ,\ee
where we have used the expansions ${\rm J}_\nu (x)\sim
(\frac{x}{2})^\nu\frac{1}{\Gamma(\nu +1)}$ and ${\rm N}_\nu(x)\sim
 -\frac{\Gamma(\nu)}{\pi}(\frac{x}{2})^{-\nu}$
 valid for $x\ll 1$.
 Note that the value for $u_m(0)$ used in \cite{Garriga:1999yh} is $\sqrt{\frac{m}{2k}}$. Moreover,
 the instability exists in the range $0<m <\frac{0.45}{M}$,
 so the perturbed black string
 is approximated by
 \be \label{bscam}
ds^2  = e^{-2k |z|}\left[-(1-\frac{2m(v)}{r})dv^2+2dvdr +r^2
d\Omega^2\right] +dz^2\ , \ee
 where \be \label{bscam2}
 m(v)=M + \int_0^{\frac{0.45}{M}} \frac{dm}{k} M(-\frac{1}{2}+2M\Omega)e^{\Omega v}u_m(0)\
 .\ee

We should compare \eqref{bscam} and \eqref{bscam2} on the brane
with the metric of a four-dimensional evaporating black hole in
the near horizon region, which takes the form \eqref{vaidya}
\cite{bardeen:1981} with
\begin{equation}\label{vaiand}
    \frac{dm(v)}{dv} \sim - \frac{T_H^2}{\hbar} \sim -
    \frac{\hbar}{M^2} \ ,
\end{equation}
the emission taking place starting from some initial $v_0$ (due to
the fact that the black hole is created from a gravitational
collapse). Differentiating \eqref{bscam2} we get
\begin{equation}\label{derbscam2}
    \frac{dm(v)}{dv} \sim - \left(\frac{1}{k M}\right)^{3/2}e^{\Omega v}
\end{equation}
where we have used the fact that $m \sim 1/M$. Neither the powers
of $\hbar$ (using the holographic relation \eqref{holrel}) nor
those of $M$ match those in \eqref{vaiand}.

Despite this negative comparison, one should not give up hopes to
find a solution with the holographic behaviour \eqref{vaiand}.
First, it is not clear whether the boundary conditions used in
\cite{GL} correspond, in the dual theory, to a black hole created
via gravitational collapse. Also, we should not exclude the
possibility that the dual Hawking radiation cannot be seen from a
classical linear perturbation analysis in 5D, but needs a full
nonlinear treatment.

\section{Analysis of the Black String null geodesics}

We will now focus on a different type of analysis. The KK
modes/CFT modes correspondence of eqs. \eqref{grap1},
\eqref{grap12} tells us that quantum effects on the brane are
induced by gravitational waves, which, in turn, propagate along
null geodesics of the full 5D spacetime. Keeping this in mind, let
us consider the full black string
\begin{equation}\label{efv2}
ds^2=e^{-2kz}\left[-\left(1-\frac{2M}{r}\right)dv^2 + 2dvdr +
r^2d\Omega^2\right]+dz^2  \ ,
\end{equation}
here written in advanced Eddington-Finkelstein coordinates (in the
RS2 case the spacetime is cut at $z=0$ and $e^{-2kz}\to
e^{-2k|z|}$).

There are two families of null geodesics in the background
\eqref{efv2} and this can be easily seen by considering the $z $
component of the geodesics equations
\begin{equation*}
    \ddot{x^{\alpha}}+\Gamma^{\alpha}_{\beta\gamma}\dot{x^{\beta}}\dot{x^{\gamma}}=0
\end{equation*}
which reads
\begin{equation*}
    \ddot{z}-k \dot{z}^2=0,
\end{equation*}
where a dot denotes a derivative with respect to the affine
parameter $\lambda$. The first family is associated to the solution
\begin{equation}\label{ge1zeropu}
   \dot z=0 \ ,
\end{equation}
i.e.
\begin{equation}\label{ge1zero}
    z=\hbox{constant} \ ,
\end{equation}
including, in particular, the null geodesics on the brane ($z=0$)
for the RS2 black string. The second, the nontrivial one, is given
by
\begin{equation}\label{ge2}
    \dot z= \frac{dz}{d\lambda} = -\frac{1}{k\lambda} \ ,
\end{equation}
which is integrated to\footnote{In the case of the RS2 black
string, the $+$ sign refers to geodesics propagating from the bulk
towards the brane and the $-$ sign to those propagating from the
brane towards the bulk.}
\begin{equation}\label{solge2}
    e^{kz}=\pm \frac{1}{k \lambda} \ .
\end{equation}

The solution to the null geodesics equations
can be obtained by considering the first integrals of motion. We
have (consider motion in the equatorial plane $\theta=\pi/2$)
\begin{equation}\label{fi1}
e^{-2kz}\left[
-\left(1-\frac{2M}{r}\right)\dot{v}^2+2\dot{v}\dot{r}+r^2\dot
\varphi^2\right]+\dot{z}^2=0
\end{equation}
and, due to the fact that $\frac{\partial}{\partial v}$ and
$\frac{\partial}{\partial \varphi}$ are Killing vectors,
\begin{equation}\label{fi2}
    e^{-2kz}\left[-\left(1-\frac{2M}{r}\right)\dot{v}+\dot{r}\right]=-E
\end{equation}
and
\begin{equation}\label{fi3}
e^{-2kz}r^2\dot\varphi=L\ ,
\end{equation}
 where $E $ and $L$ are constants.

In the case \eqref{ge1zero} with $z=0$ (on the brane, for the RS2
black string) and $L=0$ we have the usual Schwarzschild ingoing
and outgoing radial null geodesics. For ingoing geodesics
($v=$const) we have
\begin{equation}\label{schwin}
\dot r =-E  \ ,  \qquad \dot v= 0 \ .
\end{equation}
Outgoing geodesics play a key role to determine the
position of the future apparent horizon on the brane, defined as
the surface where the radius of the two-sphere $r$ has zero
divergence ($dr / dv =0$). From \eqref{fi1} and \eqref{fi2} and
considering $\dot v \neq 0$ we have
\begin{align}\label{fin1}
    &-\left(1-\frac{2M}{r}\right)\dot{v}+2\dot{r}=0 \\
    \label{fin2}
    &-\left(1-\frac{2M}{r}\right)\dot{v}+\dot{r}=-E \ ,
\end{align}
from which we get
\begin{equation}\label{schwout}
\dot r =E  \ ,  \qquad \dot v=
\frac{2E}{\left(1-\frac{2M}{r}\right)} \
\end{equation}
and, also,
\begin{equation}\label{drdvbra}
    \frac{dr}{dv}=\frac{1}{2}\left(1-\frac{2M}{r}\right) \ .
\end{equation}
Outgoing geodesics diverge for $r>2M$ ($E>0$) and converge when
$r<2M$ ($E<0$). The particular outgoing geodesics that remain on
the future horizon
\begin{equation}\label{hoho}
r=2M \ ,
\end{equation}
for which
\begin{equation}\label{drdvbra0}
    \frac{dr}{dv}=0  \ ,
\end{equation}
are characterized by $\dot r=0$, and thus
\begin{equation}\label{eeee}
E=0\ .
\end{equation}

We will perform a similar analysis for the null geodesics of the
second family (\ref{ge2}). For $L=0$ and using (\ref{solge2}) we
get\footnote{It is not difficult to generalise our analysis to
$L\neq 0$.}
\begin{align}\label{hin1}
    &-\left(1-\frac{2M}{r}\right)\dot{v}^2+2\dot v\dot{r}=-\frac{1}{(k\lambda)^4}\ , \\
    \label{hin2}
    &-\left(1-\frac{2M}{r}\right)\dot{v}+\dot{r}=-\frac{E}{(k\lambda)^2} \
    .
\end{align}
Note that due to the nontrivial motion in the bulk ($\dot z\neq
0$) (\ref{hin1}) implies $\dot v\neq 0$. We can then divide
(\ref{hin1}) by $\dot v$ and subtract (\ref{hin2}) to get
\begin{equation}\label{cucu}
     \dot r=\frac{E}{(k\lambda)^2}-\frac{1}{(k\lambda)^4}\frac{1}{\dot v}  \ ,
\end{equation}
which substituted into (\ref{hin2}) gives a quadratic equation for
$\dot v$
\begin{equation}\label{cucu1}
  \left(1-\frac{2M}{r}\right)\dot{v}^2 -\frac{2E}{(k\lambda)^2}\dot v+\frac{1}{(k\lambda)^4}
  =0  \ .
\end{equation}
This equation implies the two behaviors
\begin{equation}\label{cucu2} \dot v
=\frac{E\pm\sqrt{E^2-(1-\frac{2M}{r})}}{(1-\frac{2M}{r})}\frac{1}{(k\lambda)^2}
\end{equation}
and, from (\ref{cucu}), \begin{equation}\label{cucu3} \dot r =\pm
\frac{\sqrt{E^2-(1-\frac{2M}{r})}}{(k\lambda)^2}\ .
\end{equation}
The equations obtained
correspond to (radial) timelike geodesics in a 4D Schwarzschild
spacetime with an affine parameter $\nu \sim \frac{1}{k^2
\lambda}$ \cite{Chamblin:1999by}.

\begin{figure}[htb]
\centering \includegraphics[angle=0,width=3.7in] {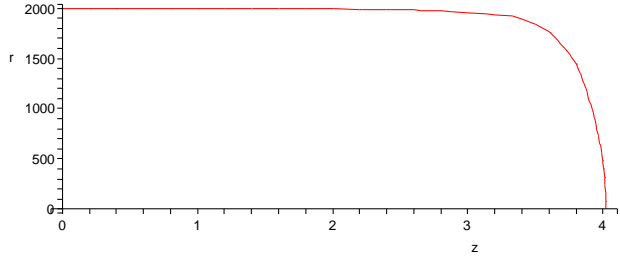}
\caption{ \label{tre}Plot of the radial geodesic $r$ as function of
$z$ (equation \eqref{rdiz}) for $M=10^3$ and $k=1$.}
\end{figure}

The purpose of this analysis is to see whether nontrivial bulk null
geodesics, presumably associated to the trajectories of the massive
KK modes, can give us some hints of quantum effects in the dual
theory on the brane. To understand if such ``quantum effects'' would
modify the horizon, eqs. \eqref{hoho}-\eqref{eeee} suggest that we
should focus on the special case $E=0$,
where we have
\begin{equation}\label{dada} \dot v
=\mp\frac{1}{\sqrt{\frac{2M}{r}-1}}\frac{1}{(k\lambda)^2}
\end{equation}
and
\begin{equation}\label{dada1} \dot r =\pm
\frac{\sqrt{\frac{2M}{r}-1}}{(k\lambda)^2}\ .
\end{equation}
Integration of  \eqref{dada1} is
straightforward and using \eqref{solge2} we get
\begin{equation}\label{rdizaa}
     \sqrt{r(2M-r)}+2M\arccos\sqrt{\frac{r}{2M}}+C =\pm \frac{e^{kz}}{k} \
     ,
\end{equation}
where $C$ is the integration constant. We shall now impose the
boundary condition that $r\to 2M$ when $z\to -\infty$.\footnote{We
will justify this choice at the end of section 5.} This implies
$C=0$ and, also, that the only real solution is the one with the $+$
sign, namely
\begin{equation}\label{rdiz}
     \sqrt{r(2M-r)}+2M\arccos\sqrt{\frac{r}{2M}}= \frac{e^{kz}}{k} \
     .
\end{equation}
The curve \eqref{rdiz} is plotted for $k=1$ and $M=10^3$ (we are
interested in the large mass regime) in fig. \ref{tre}.

From \eqref{rdiz} in the near horizon region  $r \to 2M$ we
obtain\footnote{In the case $L\neq 0$ one finds
$r_{\mathrm{bulk}}(z) \sim 2M[1-\frac{e^{2kz}}{(4Mk)^2}
(1+\frac{L^2}{4M^2})]$.
The correction to the $L=0$
case is, for large masses, very small.}
\begin{equation} \label{rpos}
r_{\mathrm{bulk}}(z) \sim 2M\left[1-\frac{e^{2kz}}{(4Mk)^2}\right] \
.
\end{equation}
It is now interesting to evaluate $dr/dv$ in this limit, and show
that is does not vanish unlike the corresponding case
\eqref{drdvbra0}. In fact from \eqref{hin2} (with $E=0$) we get
\begin{equation*}
    \frac{dr}{dv}=\frac{\dot r}{\dot v}=1-\frac{2M}{r} \sim \frac{r-2M}{2M}
\end{equation*}
and using \eqref{rpos},
\begin{equation}\label{drdv}
\frac{dr}{dv} \sim -\frac{1}{(4Mk)^2}e^{2kz} \ .
\end{equation}

\section{Hawking radiation in the holographic dual?}

Having established that massive modes ``see'' the horizon in quite
a different way, compare \eqref{rpos} with \eqref{hoho} and
\eqref{drdv} with \eqref{drdvbra0}, we shall now turn to the
possible holographic implications of our results (see also
\cite{Fabbri:2007zk}).

In the RS2 case, let us project  \eqref{drdv} on the brane to get
\begin{equation}\label{pupu}
\frac{dr}{dv} (z=0) \sim -\frac{1}{(4Mk)^2}
\end{equation}
and use the holographic relation \eqref{holrel}
\begin{equation}\label{pupu1}
\frac{dr}{dv} (z=0) \sim - \frac{\hbar N^2}{M^2} \ .
\end{equation}
We have seen in section 2 that for the case of evaporating black
holes the apparent horizon $r^\mathrm{AH}=2m(v)$ is such that, see
\eqref{vaidya} and \eqref{vaiand},
\begin{equation} \label{lumi}
    \frac{dr^\mathrm{AH}}{dv}\sim -\frac{\hbar}{M^2}\ .
\end{equation}
The similarity with \eqref{pupu1} is quite interesting, the only
difference being that in \eqref{pupu1} we have the multiplicative
factor $N^2$, indicating ``thermal emission'' due to a large number
of matter fields at the temperature $T_\mathrm{H}\sim \frac{\hbar}{
M}$.

It is not easy to justify the holographic interpretation proposed
in \eqref{pupu1} and its comparison with \eqref{lumi}. The black
string horizon is at $r=2M$, it is not receding. So why should
bulk null geodesics in the unperturbed Schwarzschild black string
know about Hawking radiation in the dual theory? We do not have a
precise answer to this question. However, we note that the
discovery of the Hawking effect \cite{Hawking:1974sw} was
performed in fixed background approximation, its implication being
that due to the quantum corrections the Schwarzschild solution
turns to a new solution with the (apparent) horizon satisfying
(\ref{lumi}).
In our case, since bulk massive modes (presumably) travel along
the null geodesics (\ref{ge2}) and they are the ones responsible
for the quantum effects on the brane, our result \eqref{pupu}
makes it reasonable to expect
that the black string too will be modified to a time-dependent
configuration with the horizon on the dual theory on the brane
evolving according to \eqref{pupu1}.

In addition to this, we know that in an evaporating black hole
scenario, due to the black hole emission apparent and event horizons
(coincident for the static Schwarzschild solution $r^\mathrm{EH}=
r^\mathrm{AH}=2M$) separate and a `quantum ergosphere' forms in
between ($r^\mathrm{EH}< r < r^\mathrm{AH}$) \cite{York:1983fx}.
Approximate calculations (valid for large $M$) of the location of
the event horizon, when the effects of the evaporation are taken
into account, show that \cite{bardeen:1981,Yobal}
\begin{equation} \label{redi}
r^{\mathrm{EH}}-2M\sim - \frac{\hbar}{ M}\ .
\end{equation}
Note that the event horizon too recedes according to \eqref{lumi}
\cite{York:1983fx}. Waves emitted from the quantum ergosphere with
wavelength  $\lambda_0 \sim r^\mathrm{AH}-r^\mathrm{EH} \sim
\frac{\hbar}{ M}$ are detected at infinity with $\lambda_{\infty}
\sim M$, which is indeed the typical size of the Hawking quanta.
As in the static case ($r^\mathrm{EH}= r^\mathrm{AH}$) there is no
emission, the existence of this region is deeply connected with
Hawking radiation.

In our case, \eqref{rpos} identifies on the brane a surface
just inside the horizon, the distance from it being
\begin{equation}\label{reldisp}
r_{\mathrm{bulk}}(0)-2M\sim -\frac{1}{Mk^2}\ .
\end{equation}
Using again the holographic relation \eqref{holrel} we get
\begin{equation}\label{reldisp1}
r_{\mathrm{bulk}}(0)-2M\sim -\frac{\hbar N^2}{M}\ ,
\end{equation}
which is very similar to \eqref{redi} except for the fact that the
``quantum ergosphere'' in the dual theory on the brane, being
multiplied by the big number $N^2$, would be much larger than in the
standard case.

\section{Features of the 5D solution dual to a 4D evaporating black hole}

The analogy between \eqref{pupu1} and \eqref{lumi} and between
\eqref{reldisp1} and  \eqref{redi} is somewhat surprising.
We shall now use it to conjecture some of the crucial features of
the actual time-dependent 5D solution allowing the holographic
interpretations \eqref{pupu1} and \eqref{reldisp1} in the dual
theory. We stress that our considerations concern configurations
where the mass is large, i.e. as long as $r_H \gg 1/k$. Indeed,
when $r_H$ becomes of order $1/k$ it is not clear whether the
holographic conjecture \cite{conj} still holds.

We recall that in braneworlds there are two different definitions
of apparent horizons, the brane apparent horizon
$r_\mathrm{brane}^\mathrm{AH}$, defined with respect to photons
which propagate along null geodesics of the 4D brane, and the bulk
apparent horizon $r_\mathrm{bulk}^\mathrm{AH}$, referring to
gravitons which follow null geodesics of the full 5D spacetime.
The static RS2 black string, for which
$r_\mathrm{brane}^\mathrm{AH}=r_\mathrm{bulk}^\mathrm{AH}=2M$, is
a special case. Indeed, in \cite{Shiromizu:2000pg} it was
numerically shown that in time-dependent braneworld black hole
solutions brane and bulk horizons are generally distinct, the
brane apparent horizon being always larger than the (brane
projected) bulk horizon.

The black strings null geodesics \eqref{ge1zero} and
\eqref{solge2}, along with their associated surfaces $r=2M$ and
$r_{\mathrm{bulk}}(z)$ in \eqref{rdiz}, \eqref{rpos}, allowed us
to reproduce such feature.
It is thus tempting to speculate that
$r_\mathrm{bulk}(z) $ simulates
the bulk apparent horizon $r_\mathrm{bulk}^\mathrm{AH}$ of
the actual 5D time dependent solution,
its projection on the brane playing the role of the event horizon
in the dual theory. In a way this is not too surprising, given
that (classical) gravitons cannot escape from inside
$r_\mathrm{bulk}^\mathrm{AH}$ and, similarly, the dual (quantum)
CFT modes cannot be emitted from the interior of
$r^{\mathrm{EH}}$.\footnote{Probably a careful analysis based on
the full solution will rather identify the brane projection of the
bulk event horizon ($\geq r_\mathrm{bulk}^\mathrm{AH}$) with the
dual event horizon $r^{\mathrm{EH}}$ ($\leq r^{\mathrm{AH}}$ in
semiclassical evaporating spacetimes). }

A  large ``quantum ergosphere''
$r_\mathrm{bulk}^\mathrm{AH} < r < r_\mathrm{brane}^\mathrm{AH}$
would then form on the brane. The dual Hawking radiation thus
suggests that 5D gravitational waves will be emitted from this
region into the bulk, the typical wavelength of the emitted waves
being of the order of $
r_\mathrm{brane}^\mathrm{AH}-r_\mathrm{bulk}^\mathrm{AH}
\sim\frac{1}{Mk^2}$.\footnote{Note that for large $AdS$ black holes,
the quasinormal modes have wavelength $\sim \frac{1}{k^2 r_+}$ with
$r_+$ the horizon radius \cite{Horowitz:1999jd}.} For large black
holes we have that $\frac{1}{Mk^2}\ll \frac{1}{k}$, and so the
(local) energy of the KK modes is large, corresponding to the
emission of ``heavy'' gravitons as in
\cite{Langlois:2002ke}.\footnote{Standard arguments
\cite{Emparan:2000rs} concerning the suppression of bulk radiation
due to small function overlap between the localized brane black hole
and the emitted bulk modes should not apply. We do not consider here
effects due to rotation \cite{Stojkovic:2004hp}.} These modes
correspond to the near-horizon `transplanckian' CFT modes in the
original derivation by Hawking. In our context, being the dual CFT
cutoff at energies $\sim k$ it is not clear what they correspond to,
since we would need to know its UV completion
\cite{ArkaniHamed:2000ds}.\footnote{Transplanckian effects are
expected to be suppressed, for large black holes, by some positive
power of $\frac{1}{Mk}$. We thank R. Emparan and N. Kaloper for
discussion on these points.}

\begin{figure}[htb]
\centering \includegraphics[angle=0,width=2in] {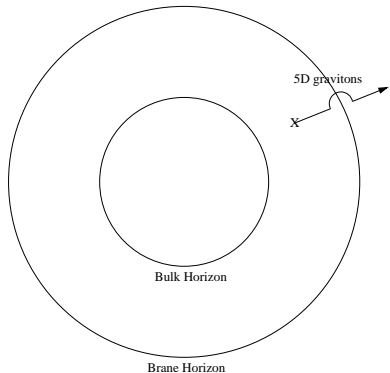}
\caption{\label{due} Dual of Hawking radiation as tunneling in
braneworld.}
\end{figure}

A suggestive possibility is the one envisaged in figure \ref{due}:
bulk massive modes emitted from the region
$r^\mathrm{AH}_\mathrm{bulk}< r < r^\mathrm{AH}_\mathrm{brane}$ will
come back to the brane just above $r^\mathrm{AH}_\mathrm{brane}$. In
the dual theory, the natural interpretation would be that of CFT
modes tunneling through the horizon, as described in \cite{tun}. It
was already suggested in \cite{lorviol} that gravitational waves
traveling between two points on the brane through a null geodesics
in the bulk can appear to travel faster than light. If so, from the
brane point of view, this would justify the ``leakage'' of the dual
CFT modes through the horizon. It remains to be seen whether such
process really takes place or not in the actual time-dependent
solution.\footnote{In a brane cosmological setting, gravitons
leaving the brane to the bulk and then bouncing back to the brane
have been shown to be present in the case of a bulk 5D Vaidya-AdS
black hole in
\cite{Langlois:2003zb}.}

Finally, we wish to show that for the full black string \eqref{efv2}
the boundary effect described in this paper disappears. Indeed, in
this case the dual CFT lives in the boundary at infinity of AdS with
no gravity. This is achieved by letting the brane position
$z_\mathrm{brane}\to -\infty$. The boundary projection of our
results \eqref{drdv} and \eqref{rpos} (or \eqref{rdiz}) gives, for
this case, $\frac{dr}{dv}\to 0$ and $r_\mathrm{bulk}\to 2M$ (i.e. no
``evaporation'' in the boundary theory and, consequently, no
``horizon splitting'' effect). This is closely related to our choice
$C=0$ made in \eqref{rdizaa} to get \eqref{rdiz}. This is a
qualitative difference with respect to the GL instability, which
always holds irrespective of whether the brane is present or not.

\section{Conclusions}

To sum up briefly the results presented in this paper,
holographic arguments applied to the propagation of bulk massive
modes in the RS2 black string \eqref{efv}, namely our results
 \eqref{pupu1} and \eqref{reldisp1}, led us to conjecture some of the
basic features of the actual time dependent solution describing an
evaporating black hole on the brane, its horizon structure
and the possible classical bulk dual of the tunneling
mechanism for Hawking radiation (fig. \ref{due}).

\section*{Acknowledgements}

\noindent G.P.P. wishes to thank Stephen Hawking for many useful
discussions. We also thank Roberto Balbinot, David Langlois
and Jos\'e Navarro-Salas
for interesting discussions. A. F. acknowledges the Spanish grant
FIS2005-05736-C03-03 and the EU Network MRTN-CT-2004-005104 for
financial support. G.P.P. is supported by PPARC and the Gates
Cambridge Trust.

\end{document}